\documentclass[11pt, oneside]{article}   
\pdfoutput=1
\usepackage{geometry}                		
\usepackage{amsmath,amssymb}
\usepackage{braket}
\usepackage{color}
\usepackage{bm}
\usepackage{graphicx}
\usepackage{wrapfig}
\usepackage{dsfont}
\usepackage[english,plain]{fancyref}

\usepackage{authblk}

\newcommand{\hc}{^{\dagger}}
\newcommand{\phc}{^{\phantom{\dagger}}}
\newcommand{\rmi}{\mathrm{i}}
\newcommand{\rmd}{\mathrm{d}}

\newcommand{\rme}{\mathrm{e}}
\newcommand{\Tr}{\mathrm{Tr}}
\newcommand{\bfi}{\mathbf{i}}
\newcommand{\bfo}{\mathbf{o}}

\begin{document}

\title{Machine learning algorithms based on generalized Gibbs ensembles}
\author{Tatjana Pu\v{s}karov\footnote{t.puskarov@uu.nl} and Axel Cort\'{e}s Cubero\footnote{a.cortescubero@uu.nl}}
\affil{Institute for Theoretical Physics, Center for Extreme Matter and Emergent Phenomena,
Utrecht University, Princetonplein 5, 3584 CC Utrecht, the Netherlands}
\date{}

\maketitle

\begin{abstract}
Machine learning algorithms often take inspiration from the established results and knowledge from statistical physics. A prototypical example is the Boltzmann machine algorithm for supervised learning, which utilizes knowledge of classical thermal partition functions and the Boltzmann distribution. Recently, a quantum version of the Boltzmann machine was introduced by Amin, {\it et. al.}, however, non-commutativity of quantum operators renders the training process by minimizing a cost function inefficient. Recent advances in the study of non-equilibrium quantum integrable systems, which never thermalize, have lead to the exploration of a wider class of statistical ensembles. These systems may be described by the so-called generalized Gibbs ensemble (GGE), which incorporates a number of ``effective temperatures". We propose that these GGEs can be successfully applied as the basis of a Boltzmann-machine--like learning algorithm, which operates by learning the optimal values of effective temperatures. We show that the GGE algorithm is an optimal quantum Boltzmann machine: it is the only quantum machine that circumvents the quantum training-process problem. We apply a simplified version of the GGE algorithm, where quantum effects are suppressed, to the classification of handwritten digits in the MNIST database. While lower error rates can be found with other state-of-the-art algorithms, we find that our algorithm reaches relatively low error rates while learning a much smaller number of parameters than would be needed in a traditional Boltzmann machine, thereby reducing computational cost.

\end{abstract}

\maketitle

\section{Introduction}
\label{sec:mnist:intro}

Supervised learning consists of approximating a function based on the input-output pairs supplied in a \textit{training} data set. Besides capturing the input-output relationships in the training set, the approximated function should be general enough so that it can be used to map new input examples supplied in a \textit{test} data set. Machine learning algorithms do this by introducing a model with flexible parameters which are learned from the training data set, often by adjusting the parameters to minimize a ``cost function'' which measures the distance of the inferred mapping from the actual one. 

One type of problem in machine learning is classification, in which the goal is to predict to which of a set of classes a sample belongs. A common approach to classification is the use of probabilistic modeling. Given some input data $\bfi$, one computes the probability $P_{\bfi}^{\bfo}$ that the input corresponds to a particular output class $\bfo$. Given an input, this produces a different probability for each possible output class, and the class with the highest probability is selected as the algorithm's prediction. The goal is to come up with probability distributions $P_{\bfi}^{\bfo}$ which have flexible parameters that can be learned from a training data set such that the predictions are closest to the correct classes.

The main goals of statistical mechanics are very similar to those of supervised machine learning. In statistical mechanics, one is interested in computing macroscopic observable quantities (such as energy density, pressure, etc.) corresponding to a given configuration of microscopic particles. One is not necessarily interested in the specific individual dynamics of each particle, since slightly different microscopic configurations lead to the same macroscopic output. This line of reasoning parallels the goals of supervised machine learning, where the input variables can correspond to a microscopic configuration of particles, and the output can be viewed as a macroscopic observable.

This conceptual connection to statistical mechanics gave rise to a popular probabilistic model based on the Boltzmann distribution, with many possible applications to machine learning \cite{hinton2006fast, hinton2006reducing, larochelle2008classification, coates2011analysis}, the so-called (restricted) Boltzmann machine (RBM/BM). In \fref{sec:mnist:classical boltzmann machines} we give a short introduction to the BM algorithm. We note here that Boltzmann machines, as well as other ML algorithms, have found interesting applications in representation of the states of quantum many-body systems~\cite{carleo2017solving}, quantum phase classification~\cite{carrasquilla2017machine,morningstar2017deep,tanaka2017detection,wetzel2017unsupervised,venderley2018phases} and classical thermodynamics~\cite{torlai2018neural}. Furthermore, in a separate relevant direction, parallels and conceptual connections between machine learning and physics have also been explored in Refs. ~\cite{mehta2014exact,koch2018mutual,lin2017why,levine2017deep}.

In this paper we develop a new learning algorithm which is inspired by recent advances in the study of the non-equilibrium statistical physics of quantum many body systems. In particular, our supervised learning model will be based on statistical ensembles that describe integrable systems, which are generalizations of the traditional Gibbs ensemble of equilibrium thermodynamics \cite{rigol2007relaxation,calabrese2012quantum2,fagotti2013reduced}. These systems are described by the so-called  generalized Gibbs ensemble (GGE), which incorporates a large number of quantities known as ``effective temperatures'', which are generalisations of the parameter $\beta=1/k_B T$.

GGEs are quantum statistical ensembles, therefore we need to study the properties of \textit{quantum} Boltzmann machines. In \fref{sec:mnist:quantum boltzmann machines} we will provide a brief introduction to these quantum algorithms, following the discussion of \cite{amin2018quantum}. We will see that it is generally not possible to efficiently train a quantum Boltzmann machine by minimizing the cross-entropy, as can be done in the classical version.

In \fref{sec:mnist:gge} we expose in some detail the structure of the GGE for a simple quantum Ising chain, which will be the basis of our model. We will show in \fref{sec:mnist:algorithm} how these effective temperatures can be treated as the adjustable parameters to be learned from the training data set, yielding a computationally cheap and reasonably effective learning algorithm. Particularly we will show that the GGE algorithm is an optimal version of the quantum Boltzmann machine, and that it is the only quantum machine where it is feasible to minimize the cross-entropy using the gradient descent technique.

We then apply a very simple version of the  GGE-based algorithm in the problem of classification of MNIST hand-written digits, and compare our results to previous established classification algorithms. This is done by representing the input data as momentum-space eigenstates of the quantum Ising Hamiltonian, and by having no hidden variables, which essentially reduces the algorithm to a classical problem, which can be easily done. Our algorithm at this point is not competitive with the high levels of precision of other state-of-the-art approaches \cite{ciregan2012multi}. The main advantage of our model instead is its simplicity, given that it manages to achieve reasonably low error rates with a comparatively very small number parameters that need to be fitted.

Our results show the simplest version of the GGE algorithm is a reasonable model for a simple treatment of the MNIST database.  At this point we do not explore further if introducing the full quantum effects, which can be done by introducing hidden variables and choosing a different basis of states, can improve the capabilities of the GGE algorithm. 

We also point out that at this point we explore only applications of the GGE algorithm to study data sets concerning classical data sources. While quantum algorithms often can provide a more efficient way to to approach classical computations (as seen for some simple examples in \cite{amin2018quantum}), a deeper potential of quantum algorithms is their suitability to study problems with truly quantum sources of data, such as the problems that have been studied through {\it classical} algorithms in ~\cite{carleo2017solving,carrasquilla2017machine,morningstar2017deep,tanaka2017detection,wetzel2017unsupervised}. Such applications to quantum data sets are beyond the scope of this paper.

\section{Classical Boltzmann machines}
\label{sec:mnist:classical boltzmann machines}

Boltzmann machines are generative models which learn the probability distribution of data using a set of hidden variables. In a typical Boltzmann machine the variables are distributed in two connected layers -- a visible and a hidden layer. Since fully connected Boltzmann machines have proved impractical, the connections between the variables are usually constrained -- in the case of a restricted Boltzmann machine the only non-vanishing weights are the interlayer ones connecting the visible and the hidden layer, while the intralayer ones are set to zero.

The idea is that the probabilities $p(v^\bfi,h)$ of configurations $(v^\bfi,h)$ can be computed, as is done in statistical physics, from a Boltzmann distribution, given by
\begin{equation}
	p(v^\bfi)=\sum_h\frac{\rme^{-\beta E(v^\bfi,h)}}{Z}, \quad  Z=\sum_{v^\bfi,h} \rme^{-\beta E(v^\bfi,h)},
\end{equation}
where $v^\bfi$ are the visible units which are fed the input data $\bfi$ and $h$ are the hidden units. The parameter $\beta$ corresponds to the inverse temperature and is traditionally set to $\beta=1$, although there have recently been explorations of temperature-based BM \cite{li2016temperature, passos2017temperature}, and $E(v^\bfi,h)$ can be thought of as the energy of a given configuration. One useful form of the energy function that can be used is that corresponding to an Ising-like model \cite{hopfield1982neural, hinton1983optimal, lieb1961two}
\begin{equation}
	\label{eq:mnist:classicalising}
	E (v^\bfi,h) = \sum_i b_i^h h_i\phc + \sum_i b_i^\bfi v^\bfi_i+ \sum_{i,j} w_{i,j}^\bfi h_i\phc v^\bfi_j,
\end{equation}
where $h_i$ and $v_j^\bfi$ are stochastic binary units which represent the hidden and the input variables, and biases $b_i$ and weights $w_{i,j}$ are free parameters that need to be learned. The goal of the algorithm is to learn, from the training data set, the optimal values of $b_i$ and $w_{i,j}$ that produce the most accurate probabilities $p(v^\bfi)$. 

The training of a classical Boltzmann machine is done by minimizing a cost function, which gives a measure of the distance between the correct probabilities $p(v^{\bfi})^{\text{data}}$ provided by the training data set and the probabilities $p(v^\bfi)$ generated by the model. A useful and commonly used cost function is the the cross-entropy, defined as
\begin{equation}
	\mathcal{L}=-\sum_\bfi p(v^{\bfi})^{\text{data}}\log p(v^\bfi).
\end{equation}
The cost function can be minimized using gradient descent. This technique consists of iterative changing of the parameters $b_i$ and $w_{i,j}$ (which we can call collectively the set of parameters $\{\theta\}$) by a small amount $\epsilon$ in the direction of steepest descent of the cost function in the parameter space. Namely, the change in a parameter $\theta$ at a given step is
\begin{equation}
	\Delta \theta=-\epsilon\,\partial_\theta\mathcal{L}.
\end{equation}
For the gradient descent technique to be useful, it has to be possible to compute the gradients $\partial_\theta\mathcal{L}$ efficiently. It can be easily shown that
\begin{equation}
	-\partial_\theta\mathcal{L} = \braket{\partial_\theta E (v^\bfi,h)} - \sum_{\bfi}p(v^{\bfi})^{\text{data}} \braket{\partial_\theta E (v^\bfi,h)}_h,
\end{equation}  
where $\braket{\dots}_h$ represents statistical Boltzmann averaging over the space of hidden variables, and $\braket{\dots}$ represents Boltzmann averaging over both hidden and visible variables and which can be efficiently approximated by a procedure called contrastive divergence~\cite{hinton2002training,hinton2012practical}. Approximately minimizing the cross-entropy, or equivalently minimizing the Kullback--Leibler divergence or maximizing the log likelihood, is then an efficient learning procedure for the classical Boltzmann machine.

The Boltzmann machine is a generative model since it learns a probability distribution which represents the training data and can, after training, generate new data with the learned distribution. It can also be used for classification of data~\cite{goodfellow2016deep}, usually as a feature extractor. While training the RBM, the network learns a different representation $h$ of the input data $v^\bfi$, which can be fed to a classification algorithm~\cite{hinton2006fast}. In this way, the RBM is used to extract the features $h$, which can then be fed to, e.g., a single softmax output layer to classify the data. Alternatively, an RBM with a part of its visible layer fed the output data $v^\bfo$ can be used to generate the joint probabilities of the inputs and the output labels. The probability of such a configuration $(v^\bfi,h, v^\bfo)$ is
\begin{equation}
	p(v^\bfi,h,v^\bfo)=\frac{\rme^{-\beta E(v^\bfi,h,v^\bfo)}}{Z}, \quad  Z=\sum_{v^\bfi,h,v^\bfo} \rme^{-\beta E(v^\bfi,h,v^\bfo)},
\end{equation}
where the energy is now
\begin{equation}
	\label{eq:mnist:classicalising2}
	E(v^\bfi,h,v^\bfo) = \sum_i b_i^h h_i + \sum_i b_i^\bfi v^\bfi_i+ \sum_i b_i^\bfo v^\bfo_i + \sum_{i,j} w_{i,j}^\bfi \,h_iv^\bfi_j + \sum_{i,j} w_{i,j}^\bfo \,h_iv^\bfo_j.
\end{equation}
The RBM then models the joint probability distribution of the input and the output data by summing over the possible states of the hidden units
\begin{equation}
	P_{\bfi}^{\bfo}=\sum_h 	p(v^\bfi,h,v^\bfo).
\end{equation} 

\section{Quantum Boltzmann machines}
\label{sec:mnist:quantum boltzmann machines}

Drawing on the success of learning algorithms based on classical statistical mechanics, the natural generalization to \textit{quantum} Boltzmann machines was recently proposed~\cite{amin2018quantum}. In this case the energy function is promoted to a Hamiltonian operator, and the data is represented as a particular quantum state. A quantum Ising Hamiltonian can be defined by promoting the classical spin variables $z_i$ to spin operators, which can be expressed in the basis of the Pauli matrices $\sigma_i^{x/y/z}$. One example considered in \cite{amin2018quantum} is the Hamiltonian corresponding to an inhomogeneous Ising chain in a transverse and a longitudinal magnetic field
\begin{equation}
	\label{eq:mnist:inhomogeneousTFI}
	H^\bfo=-\sum_i \Gamma_i^{\phantom{x}} \sigma_i^x - \sum_i b_i^{\phantom{x}} \sigma_i^z-\sum_{i,j} w_{i,j}^{\phantom{x}} \,\sigma_i^z\sigma_j^z.
\end{equation}
The superscript $\bfo$ refers to the output class of the input data. In using the quantum Boltzmann machine as a generative model  we construct separate Hamiltonians with optimal parameters for each separate output class. In this case, the Hilbert space corresponding to the Hamiltonian \eqref{eq:mnist:inhomogeneousTFI} can be split into spin variables corresponding to visible and hidden units, given by the tensor product $\ket{v(\bfi)}\otimes\ket{h}$. In the discriminative case, a part of the visible units can be reserved for the output and the Hilbert space can be split as $\ket{v(\bfi)}\otimes\ket{v(\bfo)}\otimes\ket{h}$. In either case, an input sample $\bfi$ is represented as a particular state on the visible sector of the Hilbert space $\ket{v(\bfi)}$, which corresponds to a configuration of spins in the quantum Ising chain, while the configuration of spins corresponding to hidden variables is not specified. The probabilities can now be written in terms of the quantum version of the Boltzmann distribution, also known as the canonical Gibbs ensemble, by computing the quantum average of the projection operator
\begin{equation}
	\label{eq:mnist:projection}
	\Lambda_{\bfi}=\ket{v(\bfi)} \bra{v(\bfi)} \otimes \mathcal{I}_h,
\end{equation}
where $\mathcal{I}_h$ is the identity operator acting on the space of hidden variables. The probabilities then are
\begin{equation}
	\label{eq:mnist:quantumboltzmann}
	P_{\bfi}^{\bfo} = \braket{\Lambda_{\bfi}}_{\mathrm{ GE}}^\bfo \equiv \frac{\Tr\{ \Lambda_{\bfi} \rme^{-\beta H^\bfo} \}}{Z^\bfo}, \quad  Z^\bfo=\Tr\left\{\rme^{-\beta H^{\bfo}}\right\},
\end{equation}
where the trace is over all the possible spin configurations, and $\braket{\dots}_{\mathrm{GE}}^\bfo$ denotes the quantum average of an operator using the Gibbs ensemble. 

One can attempt to train the quantum Boltzmann machine by minimizing the cross-entropy via gradient descent, as is done for the classical Boltzmann machine. It was, however, shown in Ref \cite{amin2018quantum} that this process cannot be done efficiently, since the gradient of the cross-entropy cannot be expressed in terms of the standard quantum averages, $\braket{\dots}_{\mathrm{GE}}$. The cross-entropy is given by
\begin{equation}
	\mathcal{L}=-\sum_{\bfi,\bfo} P_{\bfi}^{\bfo \, \mathrm{data}} \log \braket{\Lambda_{\bfi}\phc}_{\mathrm{GE}}^\bfo.
\end{equation}
We can then compute the gradient $\partial_\theta \mathcal{L}$, where $\theta$ stands for the parameters $\Gamma_i$, $b_i$ and $w_{i,j}$, to be
\begin{equation}
	\label{eq:mnist:quantumgradient}
	\partial_\theta\mathcal{L} = \sum_{\bfi,\bfo} P_{\bfi}^{\bfo \, \mathrm{ data}} \left( \frac{\Tr \left[\Lambda_{\bfi} \partial_\theta \rme^{-\beta H^\bfo}\right]}{\mathrm{ Tr}\left[\Lambda_{\bfi} \rme^{-\beta H^\bfo}\right]} - \frac{\Tr\left[\partial_\theta\rme^{-\beta H^\bfo}\right]}{\mathrm{ Tr}\left[ \rme^{-\beta H^\bfo}\right]}\right).
\end{equation}
It can be shown \cite{amin2018quantum} that the second term in \fref{eq:mnist:quantumgradient} can be simplified to
\begin{equation}
	\frac{\Tr \left[\partial_\theta \rme^{-\beta H^\bfo} \right]}{\Tr \left[ \rme^{-\beta H^\bfo}\right]}=\braket{\beta\partial_\theta H^\bfo}_\mathrm{ GE}^\bfo,
\end{equation}
which is a standard quantum average that can be efficiently estimated by sampling. The problem lies in the first term of \fref{eq:mnist:quantumgradient} which cannot be written as a GE average, as a consequence of the noncommutativity of the operators $[H^\bfo,\partial_\theta H^\bfo]\neq 0$. This term can be written as \cite{amin2018quantum}
\begin{equation}
	\frac{\Tr \left[\Lambda_{\bfi}\partial_\theta \rme^{-\beta H^\bfo}\right]}{\Tr \left[\Lambda_{\bfi} \rme^{-\beta H^\bfo}\right]} = -\int_0^1 \rmd t \frac{\Tr \left[\Lambda_{\bfi} \rme^{-t\beta H^\bfo} \partial_\theta H \rme^{-(1-t)\beta H^\bfo}\right]}{\Tr \left[\Lambda_{\bfi} \rme^{-\beta H^\bfo}\right]},
\end{equation}
which cannot be efficiently estimated via sampling, and makes the quantum Boltzmann machine inefficient for large systems.

An alternative approach to training the quantum algorithm was proposed in \cite{amin2018quantum}, consisting on placing an upper bound on the cross-entropy, instead of computing the absolute minimum. The upper bound is a quantity that can be efficiently estimated by sampling and thus minimized. It was shown that this bound-based approach worked well enough for some simple data sets. The problem of efficiently truly minimizing the cross-entropy of a quantum Boltzmann machine was, however, left unsolved in \cite{amin2018quantum}. 

An alternate training procedure for quantum Boltzmann machines, which can be called \textit{state based training}, was proposed in \cite{kieferova2016tomography} for training data sets that can be expressed as a density matrix $\rho^{\bfo \,\mathrm{data}}$, which allows for the source of the training data also being of quantum nature. The training data set we have discussed so far, consisting of the set of operators $\Lambda_{\bfi}$ and their associated probabilities $P_{\bfi}^{\bfo \, \mathrm{data}}$ can also be expressed as a quantum density matrix as
\begin{equation}
	\rho^{\bfo\,\mathrm{data}}=\sum_{\bfi} P_{\bfi}^{\bfo\,\mathrm{data}} \Lambda_{\bfi}\phc,
\end{equation}
although more general density matrices $\rho^{\bfo\,\mathrm{ data}}$ can be considered as well.

The goal in state-based training is then to learn the parameters in the Hamiltonian $H^\bfo$ such that the density matrix $\rho^\bfo=\exp(-\beta H^{\bfo})$ approximates the training density matrix $\rho^{\bfo\,\mathrm{ data}}$. In this case one needs to minimize some measure of the distance between the two density matrix. A convenient such measure that can take the role of a cost function, is given by the \textit{relative entropy},
\begin{equation}
	\label{eq:mnist:relativeentropy}
	\mathcal{L} = \sum_{\bfo} S(\rho^{\bfo\,\mathrm{data}} \vert\vert \rho^{\bfo}) = \sum_{\bfo} \Tr \left[ \rho^{\bfo \, \mathrm{data}} \log \rho^{\bfo \, \mathrm{data}} \right] - \Tr \left[ \rho^{\bfo \,\mathrm{data} } \log \rho^{\bfo} \right].
\end{equation}

One can now attempt to minimize this cost function by gradient descent, for which we need to compute derivatives of it with respect to parameters $\theta$ as
\begin{equation}
	\label{eq:mnist:derivativestatebased}
	\partial_\theta\mathcal{L}= \beta \sum_\bfo \left\{ -\Tr \left[ \rho^{\bfo \, \mathrm{data}} \partial_\theta H^\bfo \right] + \Tr \left[ \rho^{\bfo} \partial_\theta H^\bfo \right] \right\}.
\end{equation}
The quantum traces in \fref{eq:mnist:derivativestatebased} are standard Gibbs-ensemble-like averages. In practice, these traces can be computed by introducing an arbitrary complete basis of orthogonal states $\{\ket{n}\}$ defined such that $\braket{m\vert n}=\delta_{mn}$.
We can then write in general
\begin{equation}
	\label{eq:mnist:traceme}
	\Tr \left[ \rho^{\bfo \, \mathrm{data}} \partial_\theta H^\bfo \right] = \sum_{n,m} \braket{n \vert \rho^{\bfo\,\mathrm{data}} \vert m} \braket{m\vert \partial_\theta H^\bfo\vert n}.
\end{equation}

For a system consisting of $L$ spin-1/2 variables, the Hilbert space is spanned by a basis of $2L$ distinct orthogonal states. It is then necessary to compute in general $4L^2$ matrix elements, $\braket{n\vert \rho^{\bfo\,\mathrm{ data}}\vert m}$ and $4L^2$ matrix elements $\braket{m\vert \partial_\theta H^\bfo\vert n}$.  For each parameter $\theta$, one needs to compute a new set of $4L^2$ elements $\braket{ m\vert \partial_\theta H^\bfo\vert n}$. A quantum Boltzmann machine in general contains $\sim L$ parameters $\theta$, such that the total number of matrix elements that need to be computed to evaluate the trace \eqref{eq:mnist:traceme}, and therefore $\partial_\theta\mathcal{L}$ for all parameters $\theta$, is $\sim L^3$. 

In the next section we will introduce our new GGE-based algorithm. We will show that this GGE algorithm is a quantum algorithm which circumvents the quantum training problem from \cite{amin2018quantum}, and its cross-entropy can be efficiently minimized via sampling. In this sense, the GGE algorithm is the  quantum machine with the optimal training process. We will also show that if one attempts to perform state-based training  on the GGE algorithm, the number of matrix elements to one needs to compute  is $\sim L^2$, instead of $\sim L^3$, even though one still has $\sim L$ free tunable $\theta$ parameters.

\section{The GGE machine as the optimal quantum Boltzmann machine}
\label{sec:mnist:gge}

The main goal of this paper is to explore the utility of applying different physics-inspired probability distributions to supervised learning, and to see if they can have any advantage over the (quantum) Boltzmann distributions.

In recent years, there has been significant progress in our understanding the dynamics of quantum many-body systems out of thermal equilibrium, where concepts such as the quantum Boltzmann distribution are not applicable. In particular, there has been a large amount of work towards understanding the non-equilibrium dynamics of one-dimensional integrable quantum systems (for a review, see \cite{polkovnikov2011colloquium, eisert2015quantum, gogolin2016equilibration}). 

Integrable quantum systems are characterized by having a large number of conserved charges (whose expectation value does not change under time evolution). We can write these conserved quantities as quantum operators $Q_n$, labelled by some integer $n$, with the property that they commute with the Hamiltonian, $[H,Q_n]=HQ_n-Q_n H=0$. Generally these charges also commute with each other, $[Q_n,Q_m]=0$, and thus can all be simultaneously diagonalized. This large number of dynamical constraints usually results in these systems being exactly solvable, and enables analytic computation of certain quantities \cite[and references therein]{mussardo2010statistical}. One important result in the study of integrable systems out of equilibrium, is the realization that even after very long times, these systems never reach a state of thermal equilibrium. It is now understood that at long times, physical observables of integrable quantum systems generally reach an equilibrium state described by a generalized Gibbs ensemble (GGE) \cite{rigol2007relaxation, barthel2008dephasing, calabrese2012quantum2}, where the probabilities associated with a given state $\ket{\psi}$ are given by 
\begin{equation}
	\label{eq:mnist:gge}
	P_{\psi} = \frac{ \bra{\psi} \rme^{-\sum_n \beta_n Q_n} \ket{\psi}}{Z}, \qquad  Z=\mathrm{Tr}\left\{ \rme^{-\sum_n \beta_n Q_n} \right\}.
\end{equation}
where $\beta_n$ can be considered to be ``effective temperatures'' corresponding to each of the higher conserved quantities of the integrable model. It is important to point out that in quantum integrable models, the conserved quantities, $Q_n$, are generally extensive and can be expressed as the sum of local operators, which are essential properties needed for \eqref{eq:mnist:gge} to be a reasonable ensemble for statistical physics \cite{vidmar2016generalized}.

One appeal of using a GGE in a machine learning algorithm is that it may be possible to store a large amount of non-trivial information in the effective temperature parameters, $\beta_n$. The effective temperature variables contain information directly related to the macroscopic quantities of a system. It can then be reasonably expected that if one learns a handful of effective temperatures corresponding to a given macroscopic output, this may carry more essential information than learning a similar number of microscopic parameters, such as the coupling between two particular input variables. We then propose that in some cases, it should be more useful to learn a set of effective temperatures, $\beta_n$, than to learn the full set of microscopic couplings $\Gamma_i$, $b_i$, $w_{i,j}$. This hypothesis is further motivated by the results of \cite{li2016temperature}, which shows how the temperature is a useful hyperparameter in a traditional Boltzmann machine; in our case, this idea is exploited by having a large number of temperature-like parameters. 

The starting point of the GGE algorithm is the Hamiltonian of an integrable quantum spin chain, and its set of conserved charges. A simple integrable spin chain is described by a  homogeneous  Hamiltonian similar to \eqref{eq:mnist:inhomogeneousTFI} where we can set $\Gamma_i=\Gamma$, $w_{i,i+1}=w$, $b_i=0$. This is the prototypical transverse-field Ising (TFI) model
\begin{equation}
	\label{eq:mnist:TFI}
	H = - w \sum_{i=1}^{L}\sigma_{i}^z\sigma_{i+1}^z+\Gamma \sum_{i=1}^{L}\sigma_{i}^x.
\end{equation}
We note that in our algorithm, $\Gamma$ and $w$ are tuned as hyperparameters. The conserved charges of this model are~\cite{fagotti2013reduced}
\begin{equation}
	\begin{split}
		Q_n^+&  = -w \sum_{i=1}^L \left(S_{i,i+n}^{xx}+S_{i,i+n-2}^{yy}\right)  -\frac{\Gamma}{w} \sum_{i=1}^L  \left(S_{i,i+n-1}^{xx}+S_{i,i+n-1}^{yy} \right), \\
		Q_n^- & = - w \sum_{i=1}^L \left(S_{i,i+n}^{xy}-S_{i,i+n-2}^{yx}\right),
	\end{split}
\end{equation}
with $S^{\alpha \beta}_{i,i+l}=\sigma_i^\alpha \left(\prod_{k=1}^{l-1} \sigma_{i+k}^z \right) \sigma_{i+l}^\beta$. It can be shown that the TFI describes a system of non-interacting fermionic quasiparticles, which makes it the simplest integrable chain with spin 1/2. A GGE machine could also be built, in principle, using other more general {\it interacting} spin chain models.

The Hilbert space can again be split into spin variable corresponding to visible and hidden nodes. The probabilities for a given input and output are then given by computing the GGE average of $\Lambda_{\bfi}$ as
\begin{equation}
	P_{\bfi}^{\bfo}=\braket{\Lambda_{\bfi}}_\mathrm{ GGE}^\bfo \equiv \frac{\Tr \{\Lambda_{\bfi}\rme^{-\sum_n\beta_n^\bfo Q_n\phc}\}}{Z^\bfo}, \quad  Z^\bfo=\Tr \left\{ \rme^{-\sum_n\beta_n^\bfo Q_n\phc}\right\},
\end{equation}
where the output dependence now has been shifted only to the effective temperature parameters $\beta_n^\bfo$. This GGE machine can also be interpreted as a standard quantum Boltzmann machine, with the effective Hamiltonian $\mathcal{H}^\bfo=\sum_n\beta_n^\bfo Q_n\phc$.

The training process consists of learning the effective temperatures that yield the most accurate probabilities. This can be done again by defining a cross-entropy as
\begin{equation}
	\mathcal{L} = -\sum_{\bfi,\bfo} P_{\bfi}^{\bfo \, \mathrm{data}} \log \braket{\Lambda_{\bfi}\phc}_\mathrm{GGE}^\bfo
\end{equation}
and attempting to minimize it by gradient descent. The gradient of the cross-entropy can be computed as in the quantum Boltzmann machine as
\begin{equation}
	\partial_\theta\mathcal{L} = \sum_{\bfi,\bfo} P_{\bfi}^{\bfo \, \mathrm{data}} \left( \frac{ \Tr \left[ \Lambda_{\bfi}\partial_\theta \rme^{-\mathcal{H}^\bfo} \right]}{\Tr \left[\Lambda_{\bfi} \rme^{-\mathcal{H}^\bfo} \right]} - \frac{\Tr \left[\partial_\theta \rme^{-\mathcal{H}^\bfo} \right]}{\Tr \left[ \rme^{-\mathcal{H}^\bfo} \right]} \right),
\end{equation}
where $\theta$ now stands for the effective temperature parameters $\beta_n^\bfo$.

The full advantage of the GGE machine as a quantum Boltzmann machine now becomes evident. It is easy to see that 
\begin{equation}
	[\partial_\theta\mathcal{H}^\bfo,\mathcal{H}^\bfo]=0,
\end{equation}
which follows from the fact that all the conserved charges commute with each other. The GGE machine, built out of a large set of mutually commuting conserved charges is then the optimal quantum Boltzmann machine, in that it is the only quantum machine that can be efficiently trained by minimizing the cross-entropy. The gradient can then be simply written as
\begin{equation}
	\partial_\theta \mathcal{L} = \sum_{\bfi,\bfo} P_{\bfi}^{\bfo\,\mathrm{data}} \left(\frac{\braket{ \Lambda_{\bfi} \partial_\theta \mathcal{H}^\bfo}_\mathrm{ GGE}^\bfo}{\braket{ \Lambda_{\bfi}}_\mathrm{ GGE}^\bfo} - \braket{ \partial_\theta\mathcal{H}^\bfo}_\mathrm{GGE}^\bfo \right),
\end{equation}
which can readily be estimated by sampling. The cross-entropy can thus be efficiently minimized using a simple gradient descent technique.

We now discuss how the GGE machine is also more efficient in the case of state-based training. We suppose the training data set is given as a density matrix $\rho^{\bfo\,\mathrm{ data}}$. The goal is then to tune the effective temperature parameters such that the density matrix $\rho^{\bfo}=\rme^{-\sum_n\beta_n^\bfo Q_n\phc}$ approximates the training data density matrix. We minimize the relative entropy \eqref{eq:mnist:relativeentropy}. The gradient of this cost function is given by
\begin{equation}
	\label{eq:mnist:gradientggestate}
	\partial_\theta \mathcal{L} = \beta \sum_\bfo \left\{ -\Tr \big[ \rho^{\bfo \, \mathrm{data}} \partial_\theta \mathcal{H}^\bfo \big] + \Tr \big[ \rho^{\bfo} \partial_\theta \mathcal{H}^\bfo \big] \right\}.
\end{equation}
Again, these quantum traces can be computed by defining the orthogonal basis of states $\{\ket{n}\}$, such that
\begin{equation}
	\Tr \left[ \rho^{\bfo \, \mathrm{data}} \partial_\theta \mathcal{H}^\bfo \right] = \sum_{n,m} \braket{ n \vert \rho^{\bfo\,\mathrm{data}} \vert m} \braket{m \vert \partial_\theta \mathcal{H}^\bfo \vert n},
\end{equation}
which is so far identical to \fref{eq:mnist:traceme}, thus, in order to compute this trace, an amount of ${\sim L^2}$ nontrivial matrix elements needs to be computed. It is also convenient to express this trace in terms of the basis of eigenstates of the operator $\partial_\theta \mathcal{H}$, which we denote as $\{\ket{n}_\mathcal{H}\}$, which diagonalize this operator such that $\,_{\mathcal{H}}\braket{m \vert \partial_\theta \mathcal{H} \vert n}_{\mathcal{H}} \equiv \left( \partial_\theta \mathcal{H} \right)_n \delta_{mn}$. We can then write the trace as
\begin{equation}
	\Tr \left[ \rho^{\bfo \, \mathrm{data}} \partial_\theta \mathcal{H}^\bfo \right] = \sum_{n,m,l} \braket{ n \vert \rho^{\bfo \, \mathrm{data}} \vert m} \left( \partial_\theta \mathcal{H} \right)_l \left(c^{\mathcal{H}}_{ml}\right)^* c^{\mathcal{H}}_{ln},
\end{equation}
where we have defined the overlaps between the states in the two different bases, $c_{mn}^\mathcal{H} \equiv \,_\mathcal{H}\braket{m \vert n}$.  The problem of computing $4L^2$ matrix elements $\braket{m\vert \partial_\theta\mathcal{H}\vert n}$ is then shifted to that of computing $4L^2$ overlaps $c_{mn}^\mathcal{H}$.

The advantage of the GGE machine over the general quantum Boltzmann machine  becomes apparent when one considers computing the trace for another parameter, $\theta^\prime$. For the general quantum Boltzmann machine, every new parameter $\theta$ generates an entire new basis of $\sim L$ eigenstates, making it necessary to compute $\sim L^3$ overlaps when computing the derivative with respect to all parameters. In the case of the GGE machine, all operators $\partial_\theta\mathcal{H}^\bfo$ are simultaneously diagonalizable, therefore, no new overlaps $c_{nm}^{\bfo\,\theta}$ need to be computed when considering a new parameter $\theta^\prime$. One then only needs to compute $\sim L^2$ parameters to perform the full computation, even though one still has $\sim L$ parameters (effective temperatures) available for training.

A limitation of the GGE machine is that it can only learn information about the {\it diagonal} elements in the eigenstate basis $\{\vert n\rangle_\mathcal{H}\}$ of the density matrix $\rho^{\mathbf{o}\,{\rm data}}$, that is, only the matrix  elements of the form $\,_\mathcal{H}\langle n\vert \rho^{\mathbf{o}\,{\rm data}}\vert n\rangle_\mathcal{H}$. The GGE machine is able to  learn about these diagonal elements much more efficiently than a generic quantum Boltzmann machine would learn about a comparable number of matrix elements. Whether these diagonal matrix carry sufficient information for the algorithm to work satisfactorily will depend on the particular problem. The basis of eigenstates $\{\vert n\rangle_\mathcal{H}\}$ depends on the Hamiltonian parameters, in this case $w$ and $\Gamma$. It is then also possible to optimize the GGE machine by selecting the Hamiltonian parameters for which the matrix $\vert m\rangle_\mathcal{H}\,_\mathcal{H}\langle m\vert \rho^{\mathbf{o}\,{\rm data}}\vert n\rangle_\mathcal{H}\,_\mathcal{H}\langle n\vert$ is the most approximately diagonal. For this reason it would be interesting in the future to implement a GGE machine based on {\it interacting} integrable quantum spin chain Hamiltonians, such as the $XXZ$ Heisenberg chain chain \cite{ilievski2015complete}, rather than free models like the TFI. This is because the basis of eigenstates depends much more strongly on the Hamiltonian parameters in interacting systems, with changes in parameters inducing wider changes to the eigenstate basis \cite{sotiriadis2012zamolodchikov}, making it possible to learn properties of more diverse classes of matrices $\rho^{\mathbf{o}\,{\rm data}}$, by choosing the optimal eigenstate basis.

At this moment we do not have a general proposal for how to  find the optimal basis. Understanding the relation between different eigenstate bases as one changes the Hamiltonian parameters in an interacting model, is a remarkably difficult problem to address analytically. This problem is very relevant to quantum quench dynamics, which consist essentially of starting with an eigenstate of a given basis, and then time-evolving it with  a different Hamiltonian. The difference between the two bases is the source of the rich (and analytically challenging to compute) non-equilibrium dynamics \cite{sotiriadis2012zamolodchikov}. At the moment we only propose the Hamiltonian parameters of the integrable model may be useful hyperparameters, which in the worst case scenario can be explored and optimized by simple trial and error.

In the next section we will show a simple application of the GGE machine towards to problem of classifying handwritten digits of the MNIST data set. We will see that the amount of information learned by this GGE machine leads to reasonably accurate results, relative to the simplicity of the algorithm.

\section{A simple GGE machine for the MNIST dataset}
\label{sec:mnist:gge machine}

A simple example of a supervised learning problem is that of the classification of handwritten digits. The Modified National Institute of Standards and Technology (MNIST) database is a large database of handwritten digits used for training and testing image processing models~\footnote{Available at http://yann.lecun.com/exdb/mnist/}. The data consists of 60000 training and 10000 testing images and their corresponding labels. The input data in this case is a set of grayscale pixels which form the image of a handwritten digit. The images are $28\times 28$ pixels with 256 grayscale values. The output are the corresponding ``class'' labels of the images, i.e. the correct digits (0-9) that the images represent. Creators of the database trained several types of algorithms and achieved test set error rates ranging from 12\% for a linear classifier to 0.8\% for a support vector machine~\cite{lecun1998gradient}. At the moment, an algorithm with a committee of 35 convolutional neural networks achieves the best performance on the test set with an error of 0.23\%~\cite{ciregan2012multi}.

We now test a very simple implementation of the GGE machine towards the problem of classifying the MNIST data set. We use the simple TFI Hamiltonian, \eqref{eq:mnist:TFI}. Because of the commutativity of the conserved charges, it is possible to write a simplified GGE algorithm, where quantum effects are suppressed, by choosing to implement the data in a convenient state basis and eliminating hidden variables. This toy model then only shows that learning effective temperatures is a viable approach towards the MNIST classification problem, but it does not test if there are any advantages to introducing quantum effects. Quantum effects can be introduced by choosing a different basis and adding hidden variables, but such an implementation is beyond the scope of this introductory paper.

The Hamiltonian \eqref{eq:mnist:TFI} describes an integrable spin chain which can be diagonalized by a Jordan--Wigner and a Bogoliubov transformation~\cite{lieb1961two} to give a free fermion model
\begin{equation}
	H = \sum_k \varepsilon_k\phc \left(\eta_k\hc \eta_k\phc - \frac{1}{2}\right),
\end{equation}
where the single particle energies are $\varepsilon_k=\sqrt{w-2\Gamma\cos{\theta_k}+\Gamma^2}$ and $\rme^{\rmi \theta_k}=(\Gamma-\rme^{\rmi k})/\sqrt{1+\Gamma^2-2\Gamma\cos{k}}$~\cite{calabrese2012quantum2}. The momenta are quantized as $k_j^{\mathrm{Even}} = \frac{2 \pi}{L} (j+\frac{1}{2})$ in the even and $k_n^{\mathrm{Odd}}=\frac{2\pi j}{L}$ in the odd sector, and $j = -\frac{L}{2}, \dots, \frac{L}{2}-1$. The even or odd sector of momentum values correspond to whether periodic or antiperiodic boundary conditions are imposed on the free fermion operator. We interpret the input data, vectors $\ket{\bfi}$ of 256 grayscale values for each pixel, as the eigenstates of the Hamiltonian \eqref{eq:mnist:TFI}.  In order to do that, we binarize the MNIST data by setting a pixel value to 0 if it is smaller than 256/2, and 1 otherwise. The states are then given in the basis of the occupation number operator of the Bogoliubov fermions $\eta_k$ -- a 0 pixel means that the corresponding fermionic excitation is not occupied and a pixel 1 that it is occupied. A state belongs to the even/odd sector if it has a total even/odd number of excitations. From a practical standpoint, binarizing the data is not necessary, but we do it here to keep in line with the physical interpretation, given that fermions satisfy the Pauli exclusion principle.

The learning part of the algorithm consists in optimizing the set of effective temperatures that reproduce the output data $\bfo$.

The probability that a configuration of pixels $\ket{v(\bfi)}$ represents a digit $\bfo$ is
\begin{equation}
	\label{eq:mnist:gge_mnist}
	P_\bfi^\bfo= \frac{\Tr \left[ \Lambda_{\bfi} \rme^{-\sum_n \beta_n^{\bfo} Q_n\phc}\right]}{Z} = \frac{\rme^{-\sum_n \beta_n^{\bfo} \bra{v(\bfi)}Q_n^{\phantom{x}}\ket{v(\bfi)}}}{Z},
\end{equation}
where the second equality follows from the fact that that $\ket{v(\bfi)}$ are orthogonal eigenstates of all the conserved charges (because in this implementation we chose to interpret the data in this convenient state basis that makes it so). From this point onward, this implementation of the GGE machine is effectively classical, having eliminating the quantum averaging with this representation.

The $Q_n$ in \fref{eq:mnist:gge_mnist} are the conserved charges of the TFI chain. In terms of the Bogoliubov fermions, they have the simple form~\cite{fagotti2013reduced}
\begin{equation}
	\begin{split}
		\label{eq:mnist:Qs}
		Q_n^+&  =  \sum_k \varepsilon_k\phc  \cos\left[nk\right] \eta_k\hc \eta_k\phc , \\
		Q_n^- & = - 2 w \sum_k \sin\left[(n+1)k\right] \eta_k\hc \eta_k\phc.
	\end{split}
\end{equation}
The even charges $Q_n^+$ are defined for $n=0,1,2, \dots, L-2$, whereas the odd charges $Q_n^-$ are defined for $n=1,2,\dots, L-2$. The training of the algorithm corresponds to learning sets of Lagrange multipliers $\beta_n^\bfo$ for each digit $\bfo$ in order to produce the appropriate probabilities in \eqref{eq:mnist:gge_mnist}, with the further simplification that we only learn the conditional probabilities
\begin{equation}
	\label{eq:mnist:gge prob}
	P_\bfi^\bfo = \frac{\bra{v(\bfi)} \rme^{-\sum_n \beta_n^{\bfo} Q_n^{\phantom{x}}} \ket{v(\bfi)}}{\sum_{\bfo=0}^{9} \bra{v(\bfi)} \rme^{-\sum_n \beta_n^{\bfo} Q_n^{\phantom{x}}} \ket{v(\bfi)}},
\end{equation}
thus circumventing the difficulty of calculating the full partition function. As is standard practice in neural networks, the algorithm additionally learns ``biases''. These are Lagrange multipliers $\beta_{\mathds{1}}^\bfo$ which corresponds to a trivial charge, the identity $Q_\mathds{1}=1$.

In this simple implementation of the GGE machine, the quantum effects are completely suppressed, and therefore it can be described by an effective {\it classical} Boltzmann machine, with energy function $E(\{z_k\})=\sum_k y_k z_k$, where $z_k$ is a classical binary value, identified with the operator $z_k=\eta_k^\dag\eta_k$, and the weights are given by
\begin{equation}
	y_k=\sum_n\left(\beta_n^{\rm even}\varepsilon_k\cos[nk]-\beta_n^{\rm odd}2w\sin[(n+1)k]\right).\nonumber
	\end{equation}	
This classical reduction of the GGE machine is equivalent to demanding that the operators $\Lambda_\bfi$ be diagonal in the basis that diagonalizes the $Q_n$ charges.

The general expectation is that one can reach a reasonable level of accuracy in classifying the images by including a subset of the conserved charges $Q_n$ and learning the corresponding parameters $\beta_n^\bfo$. On the other hand, the traditional Boltzmann machine would need to learn a larger set of parameters in order to reach the same accuracy. The GGE-based algorithm can then provide a computationally much cheaper alternative. We analyze and support this in the following section.

\begin{figure}	
	\centering
	\includegraphics[width=.85\linewidth]{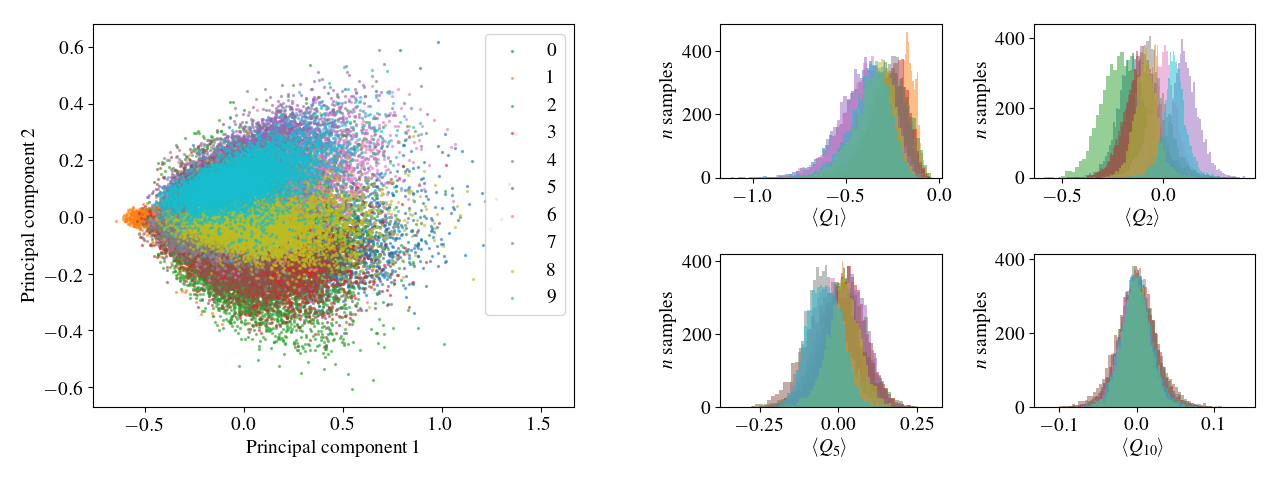}
	\caption{\label{fig:pca} Left: PCA projection onto the two first principal components of $N=21$ conserved charges calculated from the MNIST training data. PCA finds a lower-dimensional representation of the data by using an orthogonal transformation to project the data onto uncorrelated directions.
		Right: Histograms of the expectation values of several charges ($Q_1, Q_2, Q_5 \text{ and } Q_{10}$) for all the training samples. 
		In both cases, the charges were calculated using the Hamiltonian parameters $w=1.0$, $\Gamma=1.0$. The samples are colored according to the digit they represent.}
\end{figure}

\section{The algorithm and performance on the MNIST dataset}
\label{sec:mnist:algorithm}

The MNIST data is supplied as $(784,1)$ vectors, read from the $28 \times 28$ images line by line starting from the top left corner and binarized. From it, we extract a total of $N$ conserved charges which are the features we later feed to a neural network. 
For a given $N$, we calculate the expectation values of the same number of even $Q_n^+$ and odd charges $Q_n^-$ defined in \fref{eq:mnist:Qs} if $N$ is even, or one extra even charge if $N$ is odd. The features we select are therefore
\begin{equation}
	Q_n = \begin{cases}
		Q_{n/2}^+ & \text{if $n$ even}  \\
		Q_{(n+1)/2}^- & \text{if $n$ odd}
	\end{cases}, \quad n = 0, 1, 2, \dots, N-1.
\end{equation}
We stress that in the truncation $N<L$, we keep the first $N$ charges, in the sense described above. It is reasonable to question this choice, and to ask why should there be any natural preference for the charges with lower values of $n$. In physical systems, it is clear from arguments of locality \cite{pozsgay2013truncated}, that contributions from lower-$n$ charges are more important in the computation of local observables, since higher-$n$ charges are less local in coordinate space. In momentum space we see from Eq. (\ref{eq:mnist:Qs}) that the higher-$n$ charges carry information about highly oscillatory modes, while low-$n$ charges concern slowly varying modes. The argument of the importance of low-$n$ conserved charges is then akin to the usual arguments from Fourier analysis, where one needs to approximate a function in a finite interval in terms of Fourier modes, and for most functions, slowly-varying modes tend to carry fundamental information about the given function, and highly varying modes serve to further refine the approximation.


Principal component analysis (PCA) plot and histograms in fig. \ref{fig:pca} show clustering of the data for each digit indicating that the conserved charges for a particular digit do capture the similarities between different training instances of that digit. On the other hand, there are significant overlaps between the different clusters, which is expected considering that we discard a lot of information by using only $N$ charges. We use PCA only to illustrate the data, and not to change the basis before feeding the data to the network.

Having selected the features $Q_n$ with $n=0,1, \dots, N-1$, we train a fully connected single-layer neural network. The input layer $x$ consists of the $N$ selected features which are rescaled to have zero mean and unit variance, as is standard practice in neural networks \footnote{Not rescaling the input data, which is better suited to the physical analogy that we use, yields approximately $1\%$ lower validation scores as compared to the same network architecture fed with rescaled data. For example, a network with $N=51$ and $\Gamma=1.0$ yields $83.2(3)\%$ accuracy with rescaled and $81.5(4)\%$ with raw, not scaled, charges. For the comparison, each networks' hyperparameters were optimized using a random search.}.
The input layer is connected to a softmax output layer $y$ with 10 nodes, corresponding to the ten possible digits. There is an additional bias node that connects to each output node. The weights connecting the layers are contained in the matrix $W_{10\times N}$ and the bias weights in the vector $b$. The output is calculated as
\begin{equation}
	\label{eq:mnist:forward_prop}
	y = \phi(Wx+b),
\end{equation}
where $\phi$ is the activation function, in this case the softmax function~\cite{goodfellow2016deep}. This corresponds to eq.~\eqref{eq:mnist:gge prob} where, for each training sample, $x$ is a vector of the $N$ conserved charges $Q_n$ and $y$ is a vector containing the output probabilities $P^\bfo$ for all the digits $\bfo$. The weights matrix $W$ contains the Lagrange multipliers $\beta^\bfo_n$, and the biases are the multipliers corresponding to the identity charge $\beta_{\mathds{1}}^\bfo$.

The weights and biases are initialized at random values and they are learned by training the network to minimize the cross-entropy, including an $L_2$ regularization term~\cite{goodfellow2016deep, scikit-learn}.

\begin{figure}
	\begin{minipage}[c]{0.47\linewidth}
		\centering
		\includegraphics[width=\linewidth]{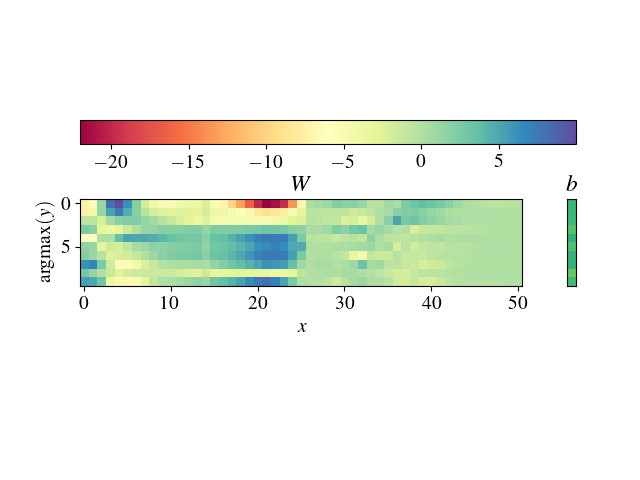}
		\caption{\label{fig:mnist:weights_biases} Typical weights $W$ and biases $b$ of eq.~\eqref{eq:mnist:forward_prop} after training a neural network described in the text. In this case, the network is fed $N=51$ conserved charges calculated from the original MNIST training data using $w=1$, $\Gamma=1$.}
	\end{minipage}
	\hfill
	\begin{minipage}[c]{0.47\linewidth}
		\centering
		\includegraphics[width=0.95\linewidth]{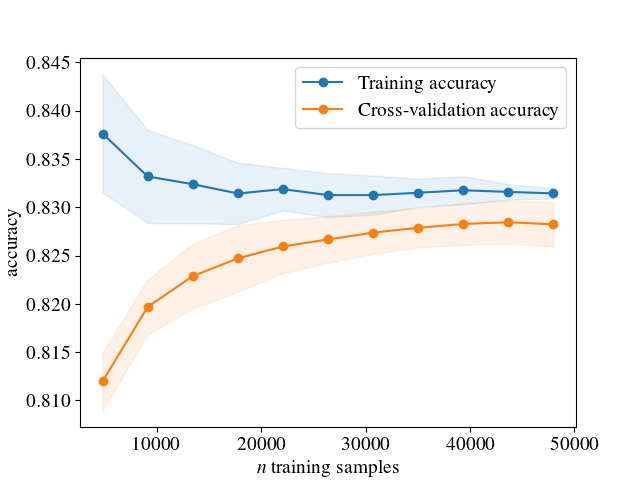}
		\caption{\label{fig:mnist:learning_curves} Learning curves for the model with the input layer size $N=51$ and the charges calculated with $w=1$, $\Gamma=2$. Learning curves are the training- and cross-validation accuracies with respect to the number of instances used for training.}
	\end{minipage}
\end{figure}

The accuracy rates depend on a number of hyperparameters (parameters of the algorithm and not the model). These include the regularization parameter, the learning rate, the stopping criterion and the maximum number of epochs~\cite{goodfellow2016deep,scikit-learn}. The optimal values of the hyperparameters are chosen using randomized parameter optimization~\cite{bergstra2012random} and stratified 10-fold cross-validation. In this procedure the full training set is split into three parts called ``folds'' (with sample distribution in each fold similar to the full dataset distribution), the model is trained on nine of the folds, and the performance is measured on the remaining, previously unseen, one. This is repeated ten times so that each fold acts as the validation set once and the performance is averaged. The procedure is repeated for each combination of the hyperparameters. The hyperparameter values for the combinations are picked at random from a specified range (for continuous parameters) or a set (for discrete parameters).

With the selected hyperparamters the network is trained on the $M=60000$ samples of the training set using stochastic average gradient descent and stratified 10-fold cross-validation. Typical weights and biases learned are shown in \fref{fig:mnist:weights_biases}, whereas \fref{fig:mnist:learning_curves} shows typical learning curves. We can conclude that the model is not overfitted, but the relatively high error might indicate a high bias with respect to the data. However, since we significantly truncate the number of input features thus losing a portion of the information, the high error is not unexpected. Furthermore, since the two curves converge, adding additional training instances would most likely not improve the performance.

\begin{figure}
	\centering
	\includegraphics[width=.45\linewidth]{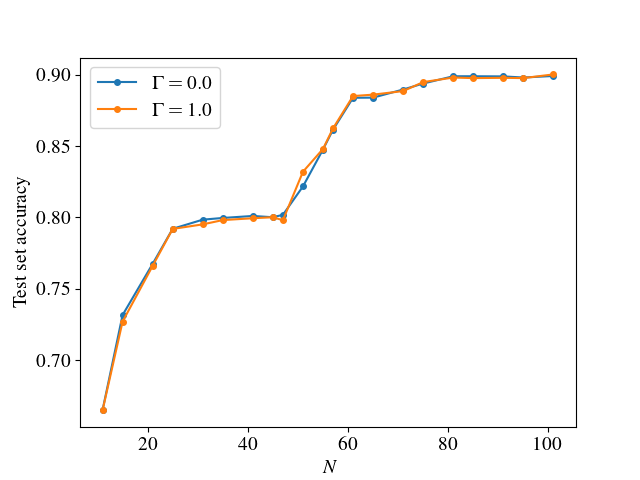}
	\quad
	\includegraphics[width=.45\linewidth]{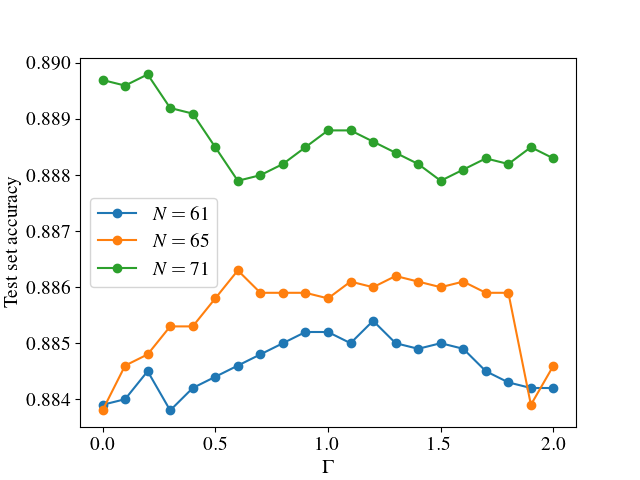}
	\caption{\label{fig:mnist:acc} The test set accuracy as a function of the size of the input layer $N$ and the Hamiltonian parameter $\Gamma$.}
\end{figure}

The algorithm performance is measured by the accuracy of the prediction on the $M_\mathrm{test} = 10000$ samples of the test set. We plot the observed accuracies with respect to the number of charges $N$ kept in the truncated GGE and with respect to the Hamiltonian parameter $\Gamma$ in \fref{fig:mnist:acc}. In this section $J=1$. As is expected, keeping a larger number of charges in the GGE improves performance. However, the accuracy rate saturates at about $90\%$ with $N \approxeq 80$ and does not improve by further adding of the charges to the ensemble - the accuracy rate when using $N=784$ charges with $\Gamma=1.0$ is still $90.9\%$. While this score is lower than what state of the art algorithms achieve, it nevertheless gives a systematic way of reducing the number of parameters of the model. The simplest neural network, trained in the original MNIST paper~\cite{lecun1998gradient}, consists of a single layer of 784 input nodes which are fed the pixel data and are connected to the 10 output nodes. This model has $n({\text{params})=n(\text{in nodes})\times n(\text{out nodes})+n(\text{bias nodes})} =  7850$ free parameters and yields $88.0\%$ accuracy. This is similar to the GGE model with $N=784$ charges at $\Gamma=1.0$ and $7850$ free parameters which yields $90.9\%$ accuracy, a slight improvement as opposed to using the raw pixel data. However, we can reach similar accuracies even with a much smaller number of parameters, as exemplified by \fref{fig:mnist:acc}. A network with $N=91$ and $\Gamma=2.0$ has $920$ parameters and yields $89.8\%$ accuracy, whereas a network with $N=61$ and $\Gamma=1.0$ has $620$ parameters and yields $88.5\%$.

For further comparison, we have trained a standard classification restricted Boltzmann machine (as defined in the introduction)~\cite{hinton2006fast} to sort the MNIST data. In this case, the network consists of a visible layer of 784 input stochastic binary nodes, a hidden layer with $n(\text{hidden nodes})$ stochastic binary nodes and an output layer with 10 binary nodes. The input, raw MNIST data rescaled to the interval $[0,1]$, is fed to the visible layer. The nodes of the hidden layer are connected to the visible nodes and are activated by a sigmoid function. The energy of this system is given in \fref{eq:mnist:classicalising}, and the weights and biases are learned using the contrastive divergence algorithm such that the energy is minimized. Using a grid search for the hyperparameters, we find that the best configuration is a network with 100 hidden nodes, thus learning $n(v^\bfi) \times n(h)+n(v^{\bfi}) + n(h)=79284$ parameters. The hidden nodes are further connected to a softmax output layer, as in the GGE case, and this part of the network is trained as a supervised model. This adds an additional $n(v^\bfo) \times n(h)+n(v^{\bfo}) =1010$ parameters, for a total of $n(\mathrm{parameter}) = 80294$. The accuracy on the test set is in this case $96.1\%$. To reiterate, in the described setup, RBM is used to extract 100 features which are used for classification. Using 100 GGE conserved quantities in contrast yields test set accuracy of $90.0\%$. This is not surprising since the features which the RBM learns have no restrictions in terms of analytic forms, which is the case with the GGE conserved quantities which are physical quantities. In order to compare the computational complexities of the two approaches, we train an RBM with $n(\text{hidden nodes})=45$ and optimized hyperparameters, which achieves a similar performance to our GGE algorithm. In this example, the RBM has a test set accuracy of $89.4\%$ with the trade off of learning a total of $36569$ parameters, while the GGE algorithm achieves $90.0\%$ with learning $1020$ parameters with $N=101$ and $\Gamma=1.0$. While there is a computational cost associated to calculating the charges we feed to the network, this is still a decrease in the total cost. The key difference here is the fact that the GGE algorithm assumes a simple Hamiltonian \eqref{eq:mnist:TFI} with homogeneous coupling, whereas the RBM learns an inhomogeneous Hamiltonian with many different coupling constants.

\section{Conclusions}
\label{sec:mnist:conclusions}

Inspired by the parallels between statistical mechanics and supervised learning, we described a machine learning algorithm based on the generalized Gibbs ensemble. This GGE algorithm turns out to be an optimal implementation of a quantum Boltzmann machine. It is the only quantum Boltzmann machine which can be efficiently trained by minimizing the cross-entropy function via gradient descent. This result follows from the fact that all the conserved charges corresponding to an integrable Hamiltonian commute with each other.

This commutativity properties also allows us to write a simplified implementation of the GGE machine. This simplified algorithm assumes that the input is an eigenstate of a simple Hamiltonian, and uses this to extract conserved charges from the inputs. Interpreting the data to be in this eigenstate basis, and eliminating hidden variables, results in all quantum effects being suppressed. This simplified version of the GGE machine can then be implemented and trained as a classical algorithm. Our numerical experiment then tests the viability of using effective temperatures as useful parameters in machine learning algorithms, but it does not yet test if there are any computational benefits to introducing quantum effects. Quantum effects can be reintroduced by working on a different basis of states and introducing hidden variables, and shall be studied in future projects. 

The effective temperatures of the GGE are directly related to macroscopic observable quantities. It is therefore expected that these parameters are more efficient quantities to learn than the typical couplings learned in a Boltzmann machine. 
The biggest advantage of our simplified classical GGE algorithm is then that it seems to achieve reasonably low error rates, while learning a comparatively low number of parameters. This advantage was shown explicitly by comparing with a restricted Boltzmann machine, where it was shown that the GGE algorithm outperforms a RBM with around 36 times the number of learned parameters. 

While our GGE algorithm currently does not outperform state-of-the art learning algorithms in terms of low error rates, it still proves to be a useful way to reduce the number of parameters to be learned. It is expected the GGE algorithm can be further improved in the future, perhaps by adding more hidden layers, and especially by the introduction of quantum effects, which might make it a more competitive alternative compared to classical algorithms.

\section*{Acknowledgments}
The authors would like to thank Dirk Schuricht for valuable discussions and for proof reading the manuscript. ACC's work  is supported by  the European Union's Horizon 2020 under the Marie Sklodowoska-Curie grant agreement 750092. This work is part of the D-ITP consortium, a program of the Netherlands Organisation for Scientific Research (NWO) that is funded by the Dutch Ministry of Education, Culture and Science (OCW). 

\bibliography{references}

\begin{thebibliography}{10}

\bibitem{hinton2006fast}
Geoffrey~E Hinton, Simon Osindero, and Yee-Whye Teh.
\newblock A fast learning algorithm for deep belief nets.
\newblock {\em Neural computation}, 18(7):1527--1554, 2006.

\bibitem{hinton2006reducing}
Geoffrey~E Hinton and Ruslan~R Salakhutdinov.
\newblock Reducing the dimensionality of data with neural networks.
\newblock {\em science}, 313(5786):504--507, 2006.

\bibitem{larochelle2008classification}
Hugo Larochelle and Yoshua Bengio.
\newblock Classification using discriminative restricted {Boltzmann} machines.
\newblock In {\em Proceedings of the 25th international conference on Machine
  learning}, pages 536--543. ACM, 2008.

\bibitem{coates2011analysis}
Adam Coates, Andrew Ng, and Honglak Lee.
\newblock An analysis of single-layer networks in unsupervised feature
  learning.
\newblock In {\em Proceedings of the fourteenth international conference on
  artificial intelligence and statistics}, pages 215--223, 2011.

\bibitem{carleo2017solving}
Giuseppe Carleo and Matthias Troyer.
\newblock Solving the quantum many-body problem with artificial neural
  networks.
\newblock {\em Science}, 355(6325):602--606, 2017.

\bibitem{carrasquilla2017machine}
Juan Carrasquilla and Roger~G Melko.
\newblock Machine learning phases of matter.
\newblock {\em Nature Physics}, 13(5):431, 2017.

\bibitem{morningstar2017deep}
Alan Morningstar and Roger~G Melko.
\newblock Deep learning the {Ising} model near criticality.
\newblock {\em arXiv preprint arXiv:1708.04622}, 2017.

\bibitem{tanaka2017detection}
Akinori Tanaka and Akio Tomiya.
\newblock Detection of phase transition via convolutional neural networks.
\newblock {\em Journal of the Physical Society of Japan}, 86(6):063001, 2017.

\bibitem{wetzel2017unsupervised}
Sebastian~J Wetzel.
\newblock Unsupervised learning of phase transitions: From principal component
  analysis to variational autoencoders.
\newblock {\em Physical Review E}, 96(2):022140, 2017.

\bibitem{venderley2018phases}
Jordan Venderley, Vedika Khemani, and E-A Kim.
\newblock Machine learning out-of-equilibrium phases of matter.
\newblock {\em Phys. Rev. Lett.}, 120.

\bibitem{torlai2018neural}
Giacomo Torlai, Guglielmo Mazzola, Juan Carrasquilla, Matthias Troyer, Roger
  Melko, and Giuseppe Carleo.
\newblock Neural-network quantum state tomography.
\newblock {\em Nature Physics}, 14(5):447, 2018.

\bibitem{mehta2014exact}
Pankaj Mehta and David~J Schwab.
\newblock An exact mapping between the variational renormalization group and
  deep learning.
\newblock {\em arXiv preprint arXiv:1410.3831}, 2014.

\bibitem{koch2018mutual}
Maciej Koch-Janusz and Zohar Ringel.
\newblock Mutual information, neural networks and the renormalization group.
\newblock {\em Nature Physics}, 14(6):578, 2018.

\bibitem{lin2017why}
Henry~W Lin, Max Tegmark, and David Rolnick.
\newblock Why does deep and cheap learning work so well?
\newblock {\em Journal of Statistical Physics}, 168(6):1223--1247, 2017.

\bibitem{levine2017deep}
Yoav Levine, David Yakira, Nadav Cohen, and Amnon Shashua.
\newblock Deep learning and quantum entanglement: {Fundamental} connections
  with implications to network design.
\newblock {\em arXiv preprint arXiv:1704.01552}, 2017.

\bibitem{rigol2007relaxation}
Marcos Rigol, Vanja Dunjko, Vladimir Yurovsky, and Maxim Olshanii.
\newblock Relaxation in a completely integrable many-body quantum system: An ab
  initio study of the dynamics of the highly excited states of 1d lattice
  hard-core bosons.
\newblock {\em Physical review letters}, 98(5):050405, 2007.

\bibitem{calabrese2012quantum2}
Pasquale Calabrese, Fabian~HL Essler, and Maurizio Fagotti.
\newblock Quantum quenches in the transverse field {Ising} chain: {II}.
  {Stationary} state properties.
\newblock {\em Journal of Statistical Mechanics: Theory and Experiment},
  2012(07):P07022, 2012.

\bibitem{fagotti2013reduced}
Maurizio Fagotti and Fabian~HL Essler.
\newblock Reduced density matrix after a quantum quench.
\newblock {\em Physical Review B}, 87(24):245107, 2013.

\bibitem{amin2018quantum}
Mohammad~H Amin, Evgeny Andriyash, Jason Rolfe, Bohdan Kulchytskyy, and Roger
  Melko.
\newblock Quantum {Boltzmann} machine.
\newblock {\em Physical Review X}, 8(2):021050, 2018.

\bibitem{ciregan2012multi}
Dan Ciregan, Ueli Meier, and J{\"u}rgen Schmidhuber.
\newblock Multi-column deep neural networks for image classification.
\newblock In {\em Computer vision and pattern recognition (CVPR), 2012 IEEE
  conference on}, pages 3642--3649. IEEE, 2012.

\bibitem{li2016temperature}
Guoqi Li, Lei Deng, Yi~Xu, Changyun Wen, Wei Wang, Jing Pei, and Luping Shi.
\newblock Temperature based restricted {Boltzmann} machines.
\newblock {\em Scientific reports}, 6:19133, 2016.

\bibitem{passos2017temperature}
Leandro~Aparecido Passos and Jo{\~a}o~Paulo Papa.
\newblock Temperature-based deep {Boltzmann} machines.
\newblock {\em Neural Processing Letters}, pages 1--13, 2017.

\bibitem{hopfield1982neural}
John~J Hopfield.
\newblock Neural networks and physical systems with emergent collective
  computational abilities.
\newblock {\em Proceedings of the national academy of sciences},
  79(8):2554--2558, 1982.

\bibitem{hinton1983optimal}
Geoffrey~E Hinton and Terrence~J Sejnowski.
\newblock Optimal perceptual inference.
\newblock In {\em Proceedings of the IEEE conference on Computer Vision and
  Pattern Recognition}, pages 448--453. Citeseer, 1983.

\bibitem{lieb1961two}
Elliott Lieb, Theodore Schultz, and Daniel Mattis.
\newblock Two soluble models of an antiferromagnetic chain.
\newblock {\em Annals of Physics}, 16(3):407--466, 1961.

\bibitem{hinton2002training}
Geoffrey~E Hinton.
\newblock Training products of experts by minimizing contrastive divergence.
\newblock {\em Neural computation}, 14(8):1771--1800, 2002.

\bibitem{hinton2012practical}
Geoffrey~E Hinton.
\newblock A practical guide to training restricted {Boltzmann} machines.
\newblock In {\em Neural networks: Tricks of the trade}, pages 599--619.
  Springer, 2012.

\bibitem{goodfellow2016deep}
Ian Goodfellow, Yoshua Bengio, Aaron Courville, and Yoshua Bengio.
\newblock {\em Deep learning}, volume~1.
\newblock MIT press Cambridge, 2016.

\bibitem{kieferova2016tomography}
Maria Kieferova and Nathan Wiebe.
\newblock Tomography and generative data modeling via quantum {Boltzmann}
  training.
\newblock {\em arXiv preprint arXiv:1612.05204}, 2016.

\bibitem{polkovnikov2011colloquium}
Anatoli Polkovnikov, Krishnendu Sengupta, Alessandro Silva, and Mukund
  Vengalattore.
\newblock Colloquium: Nonequilibrium dynamics of closed interacting quantum
  systems.
\newblock {\em Reviews of Modern Physics}, 83(3):863, 2011.

\bibitem{eisert2015quantum}
Jens Eisert, Mathis Friesdorf, and Christian Gogolin.
\newblock Quantum many-body systems out of equilibrium.
\newblock {\em Nature Physics}, 11(2):124, 2015.

\bibitem{gogolin2016equilibration}
Christian Gogolin and Jens Eisert.
\newblock Equilibration, thermalisation, and the emergence of statistical
  mechanics in closed quantum systems.
\newblock {\em Reports on Progress in Physics}, 79(5):056001, 2016.

\bibitem{mussardo2010statistical}
Giuseppe Mussardo.
\newblock {\em Statistical field theory: an introduction to exactly solved
  models in statistical physics}.
\newblock Oxford University Press, 2010.

\bibitem{barthel2008dephasing}
Thomas Barthel and Ulrich Schollw{\"o}ck.
\newblock Dephasing and the steady state in quantum many-particle systems.
\newblock {\em Physical review letters}, 100(10):100601, 2008.

\bibitem{vidmar2016generalized}
Lev Vidmar and Marcos Rigol.
\newblock Generalized {Gibbs} ensemble in integrable lattice models.
\newblock {\em Journal of Statistical Mechanics: Theory and Experiment},
  2016(6):064007, 2016.

\bibitem{ilievski2015complete}
Enej Ilievski, Jacopo De~Nardis, Bram Wouters, J-S Caux, Fabian~HL Essler, and
  Tomaz Prosen.
\newblock Complete generalized gibbs ensembles in an interacting theory.
\newblock {\em Physical review letters}, 115(15):157201, 2015.

\bibitem{sotiriadis2012zamolodchikov}
Spyros Sotiriadis, Davide Fioretto, and Giuseppe Mussardo.
\newblock Zamolodchikov--faddeev algebra and quantum quenches in integrable
  field theories.
\newblock {\em Journal of Statistical Mechanics: Theory and Experiment},
  2012(02):P02017, 2012.

\bibitem{lecun1998gradient}
Yann LeCun, L{\'e}on Bottou, Yoshua Bengio, and Patrick Haffner.
\newblock Gradient-based learning applied to document recognition.
\newblock {\em Proceedings of the IEEE}, 86(11):2278--2324, 1998.

\bibitem{pozsgay2013truncated}
Balazs Pozsgay.
\newblock Generalized gibbs ensemble for heisenberg spin chains.
\newblock {\em Journal of Statistical Mechanics: Theory and Experiment},
  2013:P07003, 2013.

\bibitem{scikit-learn}
F.~Pedregosa, G.~Varoquaux, A.~Gramfort, V.~Michel, B.~Thirion, O.~Grisel,
  M.~Blondel, P.~Prettenhofer, R.~Weiss, V.~Dubourg, J.~Vanderplas, A.~Passos,
  D.~Cournapeau, M.~Brucher, M.~Perrot, and E.~Duchesnay.
\newblock Scikit-learn: Machine learning in {P}ython.
\newblock {\em Journal of Machine Learning Research}, 12:2825--2830, 2011.

\bibitem{bergstra2012random}
James Bergstra and Yoshua Bengio.
\newblock Random search for hyper-parameter optimization.
\newblock {\em Journal of Machine Learning Research}, 13(Feb):281--305, 2012.

\end{thebibliography}
\bibliographystyle{unsrt}

\end{document}